\documentclass[12pt,preprint]{aastex}

\slugcomment{Accepted for publication by The Astrophysical Journal}

\shorttitle{Polarization of star by planet}
\shortauthors{Sengupta and Maiti}
\begin{document}

\title{POLARIZATION OF STARLIGHT BY UNRESOLVED AND OBLATE EXTRA-SOLAR
PLANET IN ELLIPTICAL ORBIT}
\author{SUJAN SENGUPTA\footnote{sujan@iiap.res.in}\and MALAY MAITI\footnote{
mith@iiap.res.in}}
\affil{ Indian Institute of Astrophysics, Koramangala, Bangalore 560 034,
India }

\begin{abstract}
We calculate the degree of linear polarization of radiation from stars having
planets that may not be resolved spatially. We assume single scattering by
water and silicate particulates in the planetary atmosphere. The dilution
of the reflected polarized radiation of the planet by the unpolarized
stellar radiation and the effect of oblateness of the planet as well as
its elliptical orbit are included.
We employ a chemical equilibrium model to estimate the number
density of water and silicate condensates and calculate the degree of
linear polarization at R band of starlight as a function of (1) mean size of
condensates, (2) planetary oblateness, (3) inclination angle, (4) phase
angle, (5) orbital eccentricity e and (6) the epoch of
periastron passage. We show that the polarization profile alters significantly
at all inclination angles when elliptical orbit is considered and the degree
of polarization peaks at the epoch of periastron passage.  We predict that
detectable amount of linear polarization may arise if the planetary atmosphere
is optically thin, the mean size of the condensates is not greater than a few
microns and the oblateness of the planet is as high as that of Jupiter.

\end{abstract}

\section{INTRODUCTION}
Polarization has always been an efficient tool to probe the physical
properties in the environment of various astrophysical objects. Recently,
it is realized that polarization could be a very important diagnostic method
for analyzing the atmosphere of sub-stellar massive object such as brown dwarfs
and extra-solar planets. \citet{sen01} predicted detectable amount of
linear polarization from L dwarfs because of the presence of condensates in the
visible atmosphere and subsequently linear polarization at optical is detected
by \citet{men02} and \citet{oso05}. The observed linear polarization can well
be explained by single dust scattering models that assumes optically thin and
rotationally oblate photosphere \citep{sen03,sen05}.

 On the other hand, the use of polarimetry in detecting and understanding
the physical properties of extra-solar planets is emphasized by
\citet{seager00,seager03, stam04}.  While \citet{seager00} modeled
the polarization due to close-in-planet or the so called roaster such as the
first discovered extra-solar planet 51 Peg b \citep{mayor95}, Stam et. al
(2004) estimated the degree of
polarization caused by planet at Jupiter's distance from the Sun. In both
the methods, the reflected flux of the planet and the Stokes vectors of the
reflected radiation are calculated by assuming circular orbit 
and blackbody radiation from the
star. The planet is considered to be spherical and the polarization is 
integrated over the illuminated portion of the disk.
 While multiple scattering polarization is calculated by \citet{stam04},
\citet{seager00} employed Monte-Carlo proceedure and computed the number of
scatterings. The later authors found that for an almost absorbing atmosphere,
higher amount of polarization may arise due to single scattering.
In both the investigations, the observable degree of polarization is estimated
by simply multiplying the flux ratio with the polarization of the radiation
from the spatially resolved planet. Since degree of polarization is a relative
measure, it does not carry any information on the radius of the circular
orbit or the planetary radius.

  However, in the most realistic situation, the planet should be oblate
because of rotation around its own axis. The oblateness of solar planets
ranges from 0.003 for Earth to 0.065 for Jupiter and 0.1 for Saturn.
Further, most of the extra-solar planets detected so far have eccentric 
orbit around the star and the eccentricity ranges from 0.0 to as high as 0.7.

  Although multiple scattering is a reasonable choice for optically thick
medium, it underestimates the amount of polarization by a few orders if
the medium is optically thin and hence if polarization is caused by single
scattering.  As mentioned before, single scattering polarization could
successfully explain the high degree of linear polarization observed from
L dwarfs.  Therefore, it's quite reasonable to investigate the amount of
polarization that could arise due to single scattering in extra-solar planets.

 In this paper, we present polarization profiles of starlight caused by
single scattering in the atmosphere of an oblate planet rotating in 
elliptical orbit and show that in certain cases the polarization may be
detectable by the existing instrumental facilities.
 
\section{SINGLE SCATTERING POLARIZATION OF STARLIGHT }

Since the dust density is assumed to be low and scatterings by atoms and
molecules (e.g., Rayleigh scattering) do
not contribute to polarization significantly, single scattering approximation
is reasonable for the region where the optical depth $\tau < 1$.
If present, multiple scattering can reduce the degree of polarization by a
few orders of magnitude \citep{sen01} because the planes
of the scattering events are randomly oriented and average each other's
contribution out from the final polarization.  Hence, the amount of observed
linear polarization could act as a probe in favour of single or multiple
scattering approximation.

In the present work, we use the formalism given by \citet{simm83} which is a 
generalization of the work by \citet{brown78}.
In this formalism, the
primary star is assumed to be a point source of unpolarized light. For most
practical application, the Stokes parameter are normalized and defined by
$[I_{refl},Q,U,V]/I_{total}$ where $I_{total}=I_{refl}+I_{star}$, $I_{refl}$
being the reflected flux or intensity from the planet and $I_{star}$ being the
unpolarized flux or intensity received from the star. We have neglected
the thermal radiation of the planet if any because for a sufficiently old
planet the thermal radiation should be much less as compared to the reflected
radiation or there may not be thermal radiation by comparatively smaller,
earth like planets. However, for young giant planets far away from the primary
star, thermal radiation should be significant.  Since $I_{refl}$ is
much less than $I_{star}$, we consider $I_{total}=I_{star}$ and hence after
we shall refer $I_{refl}$, $Q$, $U$ and $V$ to normalized Stokes parameter.
Hence the degree of linear polarization can be written as $P=\sqrt{Q^2+U^2}$

The density distribution of scatterer is calculated in co-rotating reference
frame so that it is independent of the phase angle $\Lambda$ of the planet. 
The transformation to observer frame is done by using the properties of 
spherical harmonics under rotation. In a circular orbit, the normalized Stokes
parameter $Q$ and $U$ are given as a harmonic series :
\begin{equation}
Q(k,i,\Lambda)=\sum^{\infty}_{m=0}[p_m(k,i)\cos m\Lambda+q_m(k,i)\sin m\Lambda]
\end{equation}
\begin{equation}
U(k,i,\Lambda)=\sum^{\infty}_{m=0}[u_m(k,i)\cos m\Lambda+v_m(k,i)\sin m\Lambda]
\end{equation}
where $k=2\pi/\lambda$, $\lambda$ being the wavelength, $i$ is the orbital
inclination angle. The harmonic co-efficients are given by 
\begin{equation}
\left(\begin{array}{c} p_m \\ q_m \end{array}\right)=\frac{2\pi}{k^2}
\sum^{\infty}_{l=M}F_{l2}(k)G^l_m(i)\left(\begin{array}{c} \eta_{lm} \\
\xi_{lm}\end{array}\right), m=0,1,2,3,\cdots
\end{equation}
\begin{equation}
\left(\begin{array}{c} u_m \\ v_m \end{array}\right)=\frac{2\pi}{k^2}
\sum^{\infty}_{l=M}F_{l2}(k)H^l_m(i)\left(\begin{array}{c} -\xi_{lm} \\
\eta_{lm}\end{array}\right), m=0,1,2,3,\cdots
\end{equation}
$M=max(2,m)$ and $G^l_m(i)$, $H^l_m(i)$ are given in \citet{simm83}.
$\eta_{lm}$ and $\xi_{lm}$ are related with the density distribution in the
co-rotating frame and are given by
\begin{equation}
\left(\begin{array}{c} \eta_{lm} \\ \xi_{lm}\end{array}\right)=
\alpha(l,m)\int n'(r,\theta,\phi)P^m_l(\cos\theta_i)\left(\begin{array}{c}
\cos m\phi \\ sin m\phi\end{array}\right)\sin\theta d\theta d\phi dr.
\end{equation}
where $n'(r,\theta,\phi)$ is the number density of scatterer in the co-rotating
frame, $\theta_i$ is the viewing angle,
$P^m_l$ is the associated Legendre function of the first kind and
\begin{equation}
\alpha(l,m)=\left[\frac{(2l+1)(l-m)\!}{4\pi(l+m)\!}\right]^{1/2}.
\end{equation}
At an edge-on view, $\theta_i=\pi/2$ and $\phi=0$, and hence $\xi_{lm}=0$.
$F_{l2}(k)$ is related to the scattering function and is given by
\begin{equation}
F_{lm}=\alpha(l,m)\int^1_{-1}\frac{i_1(k,\theta)-i_2(k,\theta)}{2}P^m_l(\cos\theta)d(\cos\theta).
\end{equation}
In the above equation, $\theta$ is the scattering angle, $i_1$ and $i_2$ are
the scattering functions given by \citet{van57}. $i_1$ and $i_2$ depend on
the refractive index as well as on the size and shape of the scatterer. In the
present work we consider spherical dust particles as scatterer.

Considering an ellipsoidal density distribution and using the addition theorem
of spherical harmonic,
$\eta_{lm}$ can be written as
\begin{eqnarray}\label{abc3}
\eta_{lm}=2\pi\alpha(l,m)P^m_l(0) \int^{R_1}_{R_2}n(r)dr
\int^{1}_{-1}\frac{P_l(\mu)d\mu}{[1+(A^2-1)\mu^2]^{1/2}},
\end{eqnarray}
where $R_1$ and $R_2$ are the outer and the inner equatorial axis length of the
planet, $A$ is the ratio of the length of the equatorial axis to the polar axis,
and $\mu=\cos\theta$.  We have taken multi-poles up to l=5.
We have considered up to fifth harmonic, i.e., m=0,1,2,3,4,5. However,
since the density distribution is symmetric about the orbital plane, the first,
the third and the fifth harmonics are zero and the fourth harmonic is small
as compared to the second harmonic. In other word, for the adopted density
distribution, the odd number harmonics are zero and the degree of polarization
is determined mainly by the second harmonic.

We convert $n(r)dr$ into $n(P)dP$ by using the equation of hydrostatic
equilibrium $n(r)dr=n(P)dP/g\rho(P)$,
where $P$ is the pressure at different geometrical height, $\rho$ is the mass
density at different pressure scale, and $g$ is the surface gravity (which
can be assumed to be constant for a geometrically thin atmosphere). The
degree of polarization is calculated at wavelengths ranging from 0.5486 to
0.8491 $\mu m$ and averaged by using the response function of R band Bessel
filter.

\section{EFFECT OF ORBITAL ECCENTRICITY}

The effect of orbital eccentricity enters through a multiplicative factor 
$h(\Lambda,e)$ to the co-efficient of the dominant harmonic $p_2$ and $v_2$
\citep{brown82} and it is given by
\begin{equation}
h(\Lambda,e)=\frac{[1+e\cos(\lambda-\lambda_p)]^2}{(1-e^2)^2}
\end{equation}
where $e$ is the orbital eccentricity and $\lambda$ is the true anomaly which
is related to the eccentric anomaly $E$ by the relationship
\begin{equation}
\tan\left(\frac{\lambda-\lambda_p}{2}\right)=\left(\frac{1+e}{1-e}\right)^{1/2}
\tan\frac{E}{2},
\end{equation}
and $\lambda_p$ is the longitude of the periastron.
The eccentric anomaly $E$ is related to the orbital phase angle through
Kepler's equation
\begin{equation}
 E-e\sin E=\Lambda-\Lambda_p
\end{equation}
where $\Lambda-\Lambda_p= 2\pi(t-T_0)/P$ is the mean anomaly, $t$ being
any epoch of time, $P$ is the orbital period and $T_0$ is  the epoch of
periastron passage. We have assumed no drastic seasonal change in the
planetary atmosphere and that the atmospheric T-P profile remains the same
through out the planetary year.

For a circular orbit, the  polarization profile is determined by the second
harmonic only. But for an elliptical orbit, the first and the third harmonics 
are non-zero although the second harmonic remains dominant.

\section{THE ATMOSPHERIC MODELS} 

We have adopted the atmospheric models of 
extra-solar planets given in \citet{sudars03}. The Temperature-Pressure (T-P) 
profiles are taken for models of a class II
and a class V planets or closed-in-planet also known as roaster. 
The atmosphere of class II planets should have water as condensates while
class V planets should contain silicate in their upper atmosphere.  
In calculating the T-P profile of the planet Ups And d (a class II planet),
\citet{sudars03} considered the surface gravity 
to be $2\times 10^{4}$  cms$^{-2}$ while for the planet HD 209458 b (a class V
planet) the surface gravity is taken to be 980 cms$^{-2}$. In the present work,
we have taken the same values for the surface gravity of the two representative
planets.  We have
assumed that the thermal radiation of the planet is negligible compared to
the reflected radiation so that the contribution to polarization comes only
from the reflected radiation.

The dust distribution in the atmosphere is calculated based on the one
dimensional cloud model of \citet{coo03}.
This model assumes chemical equilibrium throughout the atmosphere, and uniform
density distribution across the surface of an object at each given pressure
and temperature. Under these assumptions, the number density of cloud particles
is given by
\begin{eqnarray}\label{density}
n(P)=q_c \left( \frac{\rho}{\rho_d}\right)
\left(\frac{\mu_d}{\mu}\right)
\left(\frac{3}{4\pi r^3}\right),
\end{eqnarray}
where $\rho$ is the mass density of the surrounding gas, $r$ is the cloud
particle radius, $\rho_{d}$ is the mass density of the dust condensates,
$\mu$ and $\mu_d$ are the mean molecular
weight of atmospheric gas and condensates respectively.
The condensate mixing number ratio ($q_c$) is given as
$q_c=q_{below}P_{c,l}/P$
for heterogeneously condensing clouds where $q_{below}$ is the fraction of
condensible vapor just below the cloud base, $P_{c,l}$ is the pressure at
the condensation point, and $P$ is the gas pressure in the atmosphere.
The condensation curves for water and forsterite condensates are taken 
from \citet{sudars03}.  The values of $\mu_d$, $\rho_{d}$ and $q_{below}$
for forsterite and water are taken from \citet{coo03}.
The real part of the refractive index for forsterite is fixed at
1.65 at any wavelength and the imaginary part is taken by interpolating the
data given in \citet{sco96}. For the refractive index of water, we have used
the data given in \citet{segel81}.

Apart from the calculation of the grain number density, the location of the
cloud in the atmosphere plays an important role in determining the amount of
polarization. The location of the cloud base for different atmospheric models
and different chemical species is determined by
the intersection of the T-P profile of the atmosphere model and the
condensation curve $P_{c,l}$ as prescribed in \citet{coo03}. 
Taking the condensation curve for forsterite and water as given in 
\citet{sudars03}, we determine the base of the cloud from the T-P profiles
of Ups And d and HD209458 b calculated by the same authors.

At present, there is no convincing justification in favor of any specific
form of the particle size distribution function.  In the present work,
we adopt a log-normal size distribution function used by \citet{ack01} and
\citet{sau00} with a fixed width at 1.3.  Following \citet{sudars03}, we have
taken the mean particle radius for water 5.0 $\mu m$ and that for silicate
(forsterite) 10.0 $\mu m$. However, in order to show the effect of mean grain
size on the degree of polarization we have also presented polarization profile
as a function of mean particle size keeping the T-P profile the same as that
calculated by \citet{sudars03} for water with mean grain size 5.0 $\mu m$.

In multiple scattering, the mean particle size 
should be different at different pressure scale. However, for single
scattering, a fixed mean particle size of a particular condensate is 
sufficient \citep{sen05}.

\section{THE OBLATENESS OF PLANETS}

The rotationally induced oblateness of solar planets has been discussed
in details by \citet{hub84} and \citet{mur00}. Recently, the formalism
for oblateness is extended to extra-solar planets by \citet{bar03} who 
used Darwin-Radau relationship
\begin{equation}\label{obl}
f=\frac{\Omega^2R^3_e}{GM}\left[\frac{5}{2}\left(1-\frac{3}{2}K\right)^2+
\frac{2}{5}\right]^{-1}
\end{equation}
\citep{mur00,bar03} to relate rotation to oblateness.
In the above equation, $K=I/MR^2_e\leq2/3$ is the moment of inertia parameter
of an object with moment of inertia $I$. The Darwin-Radau relationship is exact
for uniform density objects ($K=0.4$) and provides a reasonable (within a 
few percent of errors) estimation of the oblateness of the solar planets.
At 1 bar pressure level, the oblateness $f$ of Jupiter,
Saturn and Earth are 0.065, 0.098 and 0.003 respectively. \citet{bar03} modeled
the planet HD209458b and estimated its oblateness to be about 0.00285.
Apart from rotational effects, tidal interaction of a closed in planet
with its primary star may also impose an ellipsoidal shape extending
toward the star. In the present work we ignore such effects. Moreover,
the estimation of the moment of inertia of the planets needs density
distribution, radius and mass of the planet along with its rotational
period and so it is highly model dependent. In the present work we adopt
a wide range of values for oblateness, the maximum value is equal to that of
Jupiter ($f=0.065$) and the minimum value is equal to that of Earth($f=0.003$).

\section{RESULTS AND DISCUSSION}
We present in figure 1 and in figure 2, the polarization profiles at
R band (Bessel filter) of star with a planet having
forsterite as a major condensates in its atmosphere.  Silicate 
and iron can form high in the atmosphere of class V roasters that orbit their
stars around 0.05 AU. Although presence of more than one species affects the
amount of polarization because of differences in size, optical properties, and
location, we assume forsterite as the major condensate in our model for
simplicity. However, the change in the amount of polarization by the
incorporation of other condensate species cannot be determined due to the
dominant effect of other parameters such as inclination angle, orbital 
eccentricity, oblateness etc.
Figure~3 presents the degree of polarization as a function of planetary
oblateness. Figure~4 presents the polarization profile for planets with
non-circular orbit and the change in degree of polarization caused by the
change in orbital eccentricity is shown in figure~5 .

 Our polarization profiles for the orbital eccentricity e=0 i.e., for
circular orbit, presented in figure 1 and figure 2, are qualitatively the
same as that presented by \citet{stam04}.
When the inclination angle $i=0$, the observed part of the planet that is
illuminated is constant and hence the degree of polarization is constant.
The maximum amount of polarization produced by the planet is the same for
all values of $i$. When the inclination angle $i=90^{o}$, the polarization
is zero at the fractional orbital period $t/P= 0.25$ and 0.75 that correspond
to the phase angle $\Lambda=180^{o}$ and $0^{o}$ respectively because the
planet's night and day side are turned toward the observer in these cases.
Apart from the geometrical asymmetry of any object, another asymmetry
plays important role for a binary star or a planet around a star in determining
the polarization profile. This second effect is due to the asymmetry in the
position of the object (or the scatterer) with respect to the observer and the
source of light. As the symmetry axis of the planet  changes position
with respect to the observer, the total number of effective scatterers also
changes. This effect can be visualized as the amount of light that is scattered
towards the observer from the net illuminated region.
In a  circular orbit, the observed degree of polarization is determined by the
scatterers that are present within the illuminated area facing towards the
observer. The observed degree of polarization is zero when the night side of
the object is towards the observer and it increases as the illuminated area
towards the observer increases.  The fraction of the area that is illuminated
depends on the inclination angle as well.  As the inclination angle decreases,
the orbital plane becomes more and more face on towards the observer.
Consequently, the illuminated area towards the observer increases yielding
into higher amount of polarization that can be seen by the observer.

However, the polarization profile alters drastically when the orbit is
non-circular as can be seen in figure~4. 
The degree of polarization peaks when the planet passes its periastron.
Hence, the degree of polarization is time variable even if the inclination
angle is zero degree.  The change in the amount of polarization is determined
by the eccentricity of the orbit as can be seen from Figure 5.

As discussed by Brown et al (1982), the "weighted"
or effective optical depth (as seen by the source) of the scattering region
idealized as point scatterer containing N particles is proprtional to the
square of the ratio between the semi-major axis $a$ and the stellar separation
at longitude $\lambda$ given by $R(\lambda)=a(1-e^2)/[1+e\cos(\lambda-
\lambda_p)]$.  For a circular orbit this ratio is identically 1 but for
elliptical orbit the ratio is $[1+e\cos(\lambda-\lambda_p)]/(1-e^2)$.
This means, for an elliptical orbit the effective optical depth of the
scattering region as seen by the source is increased by a factor of
$h(\lambda,e)$ given in section~3 .
For a circular orbit, e=0 and the effective optical depth and hence the
degree of polarization becomes independent of the longitude. When
$\lambda=\lambda_p$, i.e., when the planet is at the periastron, the effective
optical depth as seen by the source is the maximum yielding maxiumum
amount of polarization.

As shown by \citet{sen01,sen03,sen05}, the oblateness of the
atmosphere plays a crucial role in determining 
the degree of polarization.
The net degree of polarization is calculated by  integrating the polarization
at all points over the planetary disk. Since, we take the number  density of
scatterers distributed about some symmetry axis, the
positive part of polarization gets cancelled out by the negative part . Now as
oblateness increases, the net cancellation decreases owing to the departure
from spherical symmetry.  Consequently, the disc
integrated polarization would vanish if the geometry is perfectly spherical
and departure from sphericity would give rise to non-zero polarization. An
increase in oblateness means more asymmetry and hence more net non-zero
polarization.  Figure~1 shows that if the
oblateness of the planet is as small as 0.003 (similar to that of Earth) then
the degree of polarization is of the order of $10^{-4}$ for the mean grain
size of forsterite $10.0$ $\mu m$. The degree of polarization estimated
by \citet{seager00} for similar planets is of the order $10^{-6}$ for grain
size 1-10 $\mu m$.
This is because of the fact that multiple scattering reduces the amount
of polarization by a few orders of magnitude as discussed in section~2.
However, if the oblateness of the object is increased, 
the degree of polarization increases substantially as shown
in figure~2  and could be at the range of present detectibility. In figure~3
we present the degree of polarization by forsterite grain of mean radius
10 $\mu m$ for different values of the planetary oblateness.

  Figure 6 shows the polarization profiles of star
with class II planet that orbits at a distance of 1-2 A.U and hence should have
atmospheric temperature about 250 K. This type of planet should have a 
tropospheric cloud layer of water along with lesser amount of methane and 
ammonia.

The amount of polarization is although dependent on the optical property
of the condensates, the size of the condensates play the dominant role
in determining the amount of polarization. According to the adopted
chemical equilibrium model by \citet{coo03}, the particle number density
increases with the decrease in grain size. As a result, the amount of 
polarization increases substantially when the mean particle size is
decreased.  Therefore, high degree of polarization might arise in a
less oblate planet if the grain size is sufficiently small. On the other hand,
large amount of polarization can be produced if the grain size is
large but the oblateness is high as shown in figure 6. In figure~7, we present
the degree of polarization caused by water particulates with different
mean size. However, in order to understand the effect of grain size, we have
kept the T-P profile the same as that calculated by \citet{sudars03} with
mean water particulate size 5 $\mu m$. 

It may not be possible to infer from the polarization
profile, the specific nature of the condensates, the oblateness of the planet
or the orbital inclination angle as the amount of polarization would arise
by a combination of all these parameters. However, if the distance of the
planet from the primary star is known, then the nature of the condensate
can be determined. Exact understanding on the chemical equilibrium may provide
the size of the condensates. As a result, a combination of the oblateness
and the inclination angle of the planet can be determined from the polarization
data which in turn can provide information on the mass and the rotational
velocity of the planet. However, the periastron angle and the orbital
eccentricity can be inferred, irrespective of the amount of polarization, from
the phase angles at which the maximum and the minimum amount of polarization
arise.

\section{CONCLUSION}
The important message that is conveyed in this paper is that single scattering
in a rotationally induced oblate planet may cause sufficient amount of
linear polarization that could be detected by using the available
instrumental facilities. The amount of linear polarization would be
sufficiently large if the planet is highly oblate and the mean size of the
condensates is not greater than a few micron. If detected, the polarization
profile will provide information on the oblateness and orbital inclination
of the planet and hence on the spin period and mass of the planet. Further,
the time dependent polarization profile would provide the orbital
period and the eccentricity. Our present discussion is confined to stars
with single planet. The case for more than one planet orbiting a star at 
different inclination and phase angle
is worth investigating as that should change the amount and the periodic
nature of the variability of the polarization. The prsent investigation
is aimed for the polarimetry of already known extra-solar planets. Since,
the orbital separation and the emitted flux from the parent star are known,
one can employ the theoretical inferences about the composition and location
of the condensates in the planetary atmosphere. However, in general, the
degree of polarization should depend on the evolution of the planets.
\citet{baraffe03} have discussed on the time evolution of extra-solar
planets and showed how the radius and hence the surface gravity as well as
the effective temperature of an irradiated planet changes with time. Since
the formation, chemical evolution and the geometrical location of condensates
change with the change in surface gravity and atmospheric temperature profile,
the degree of polarization, wheather by single scattering or by multiple 
scattering, would also alter with time. Therefore, time evolution of the
polarization profile is worth investigating.

\acknowledgments

We are thankful to the referee for several constructive comments and
useful suggestions that have improved the clarity and the quality of the
paper.This research has made use of NASA's Astrophysics Data System.

\clearpage

\plotone{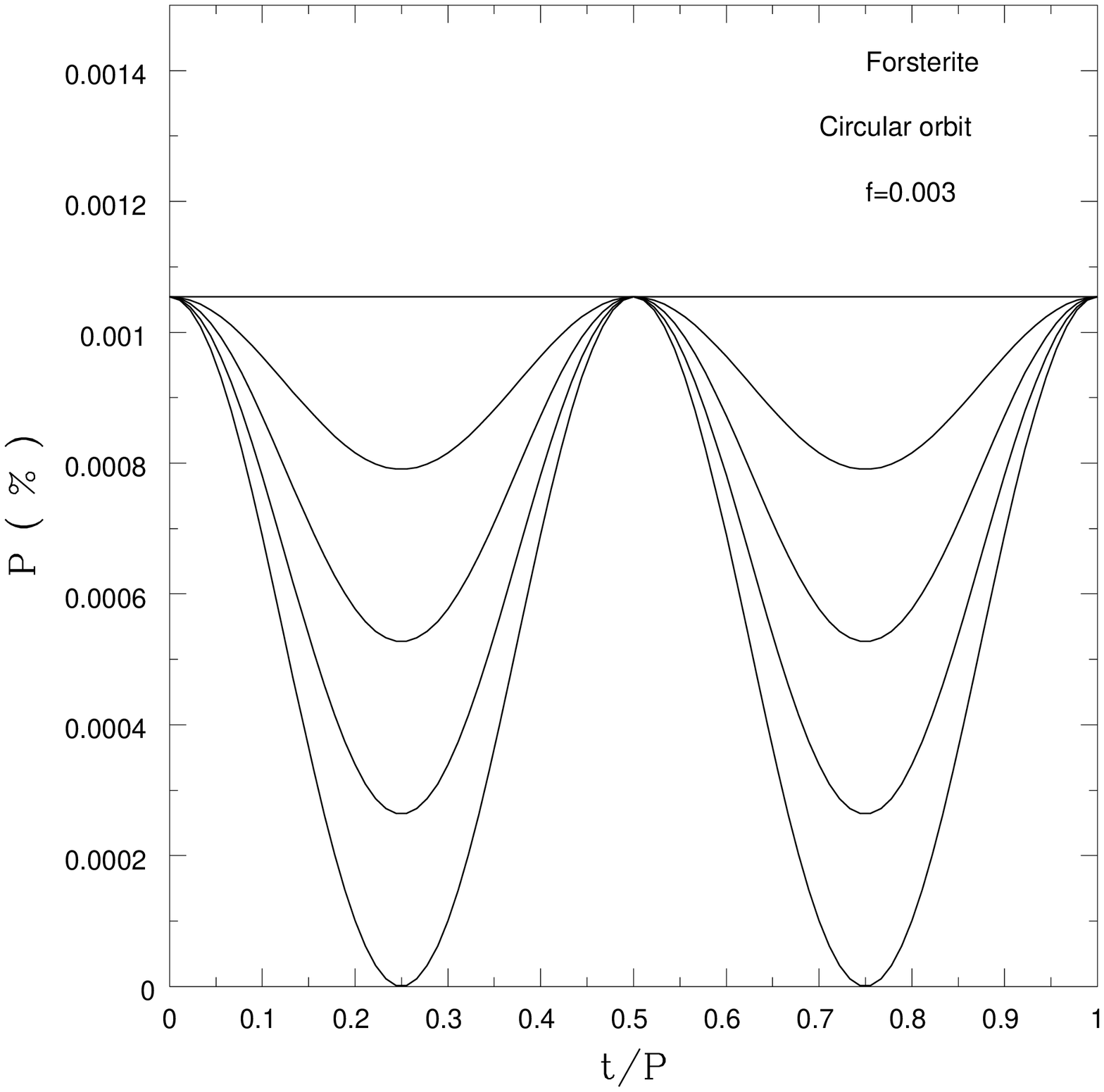}
\figcaption[f1.ps]{Degree of linear polarization at Bessel R band  
caused by forsterite grain with mean radius $r_0=10 \mu m$. The polarization
is plotted as a function of the fractional orbital period $t/P$ where $t$ is
time and $P$ is the orbital period.
Models for planet with oblateness f=0.003 and circular orbit.
From bottom to top at $t/P=0.25$, the curves represent polarization with 
inclination angles $i=90^o$, $60^o$, $45^o$, $30^o$ and $0^o$ respectively.
\label{figure1}}

\plotone{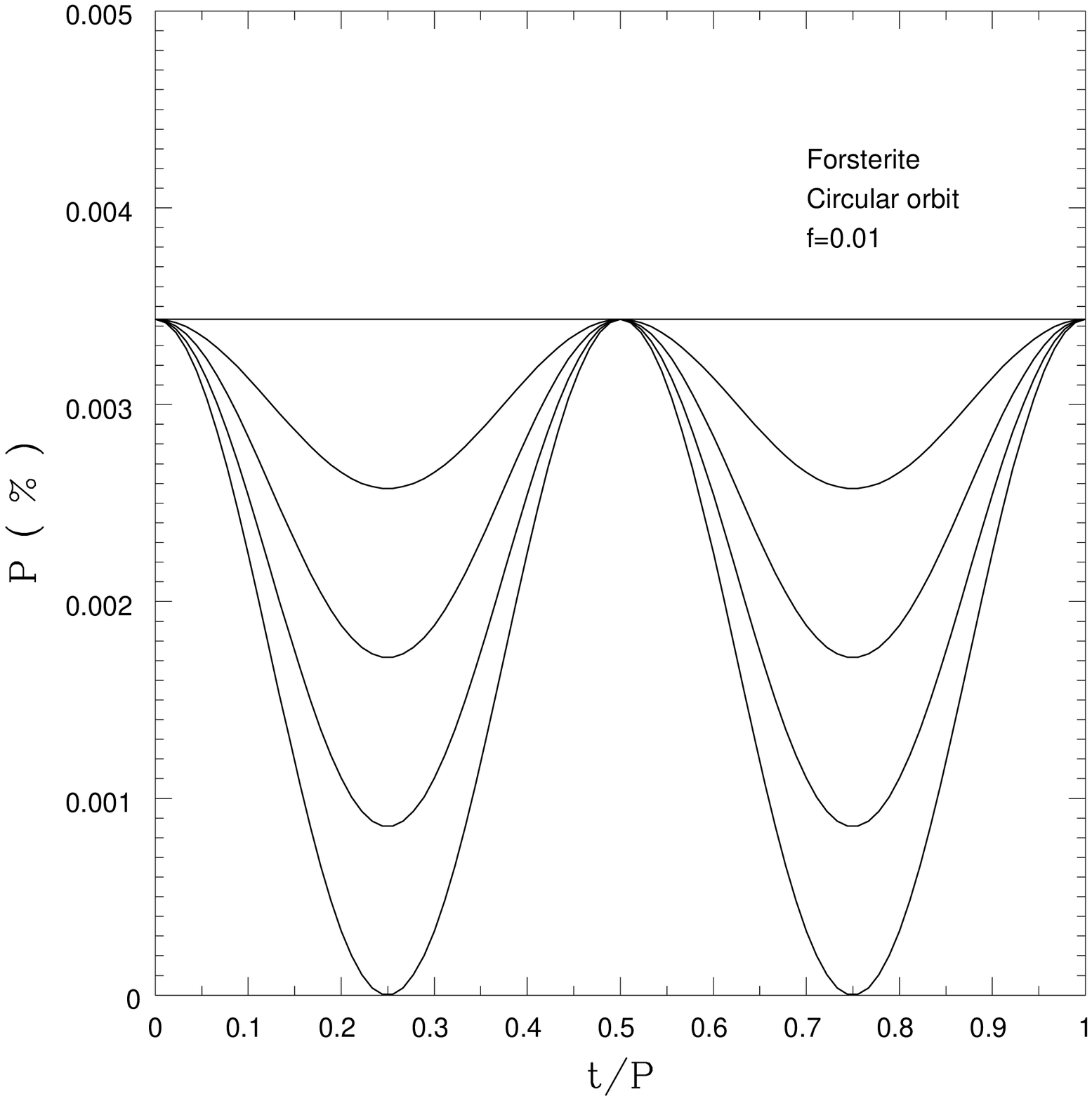}
\figcaption[f2.ps]{Same as figure 1 but for planet with oblateness f=0.01.
\label{figure2}}

\plotone{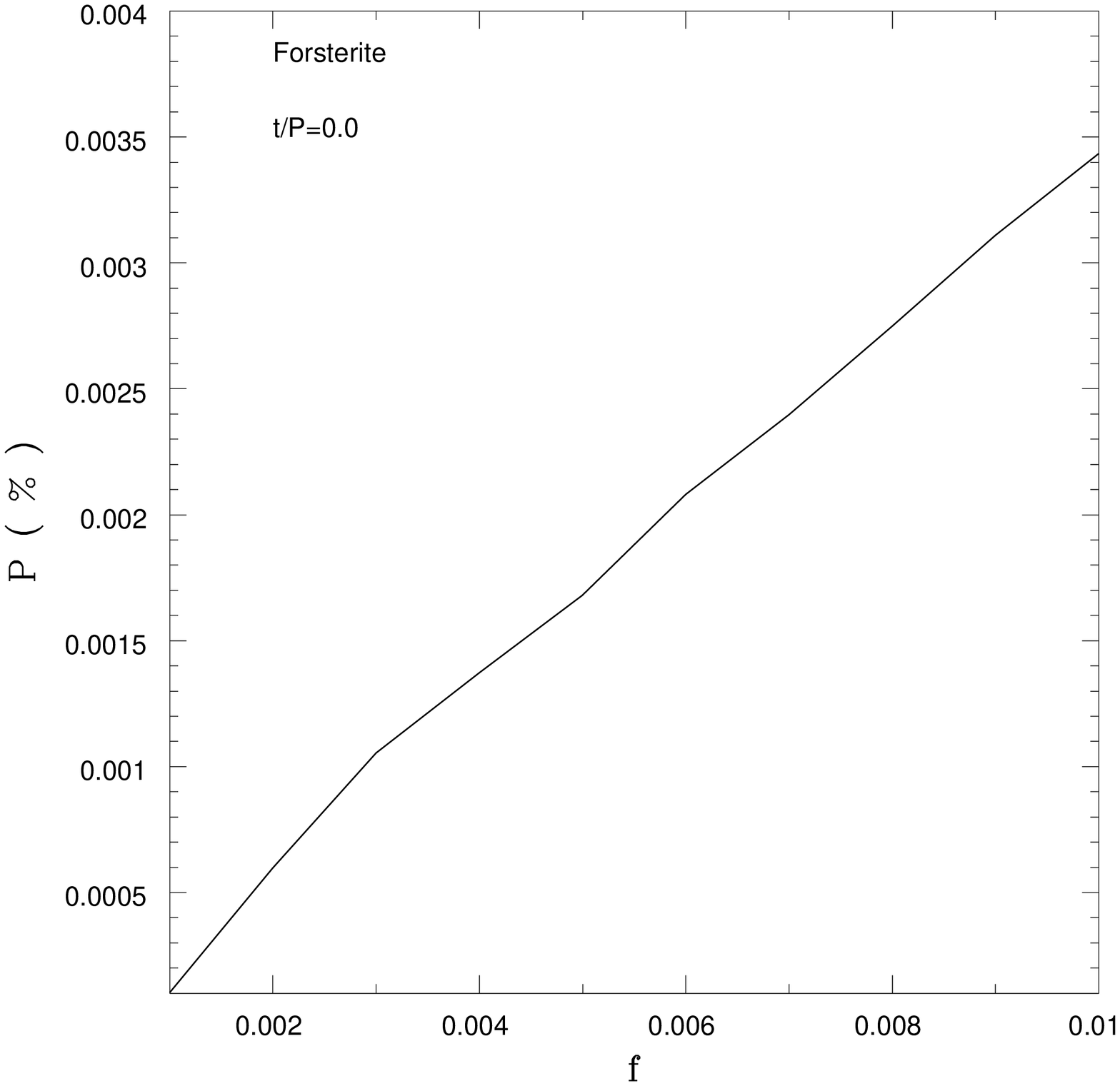}
\figcaption[f3.ps]{Degree of polarization at Bessel R band
as a function of oblateness f.  Circular orbit with inclination angle $i=90^o$.
The linear polarization by planet with forsterite with $r_0=10$$\mu m$
is calculated at the fractional orbital period $t/P=0.0$.
\label{figure3}}

\plotone{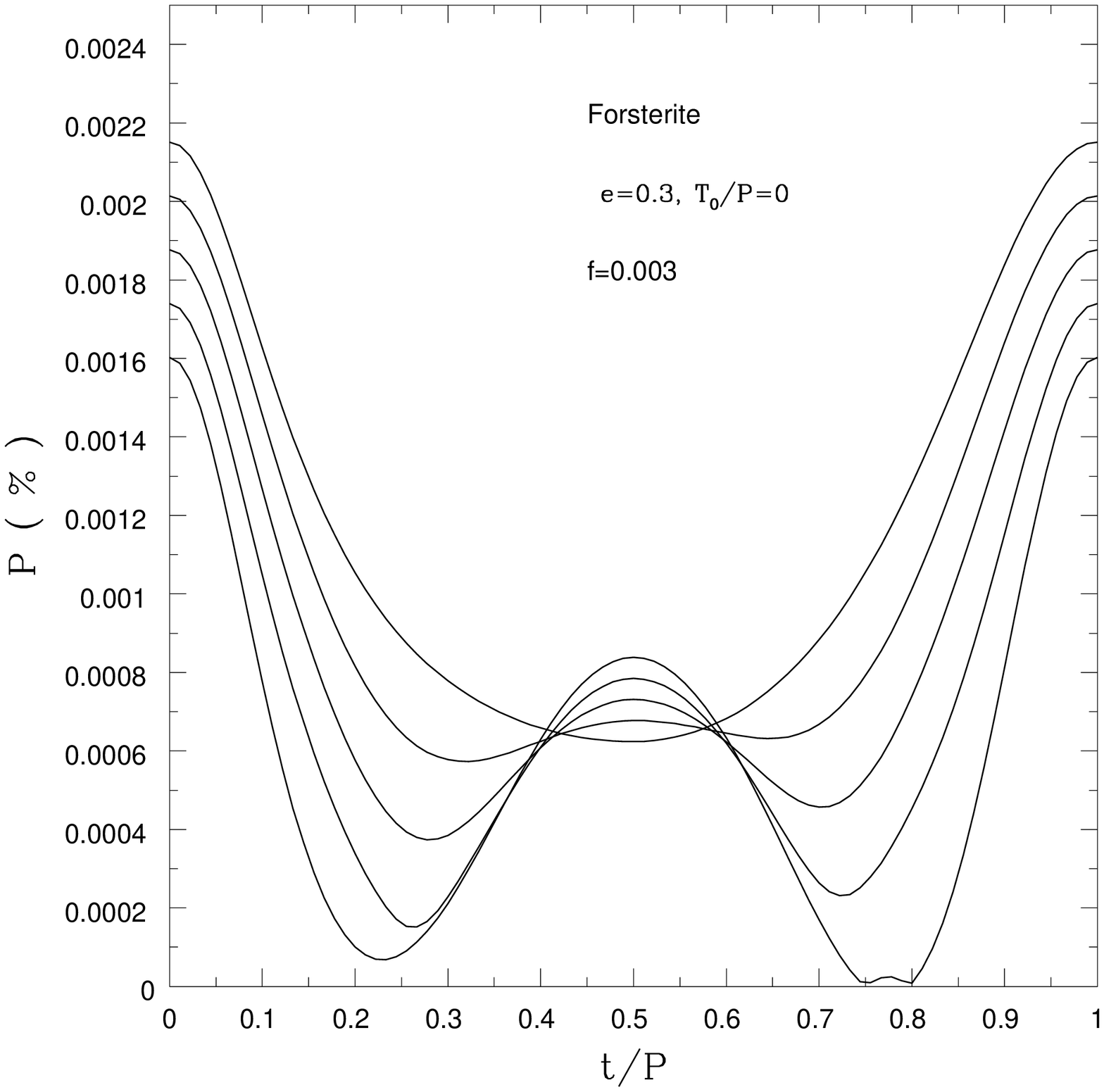}
\figcaption[f4.ps]{Same as figure 1 but for planets orbiting in elliptical
orbit with eccentricity $e=0.3$ and the epoch of periastron passage
$T_0/P=0$.
\label{figure4}}

\plotone{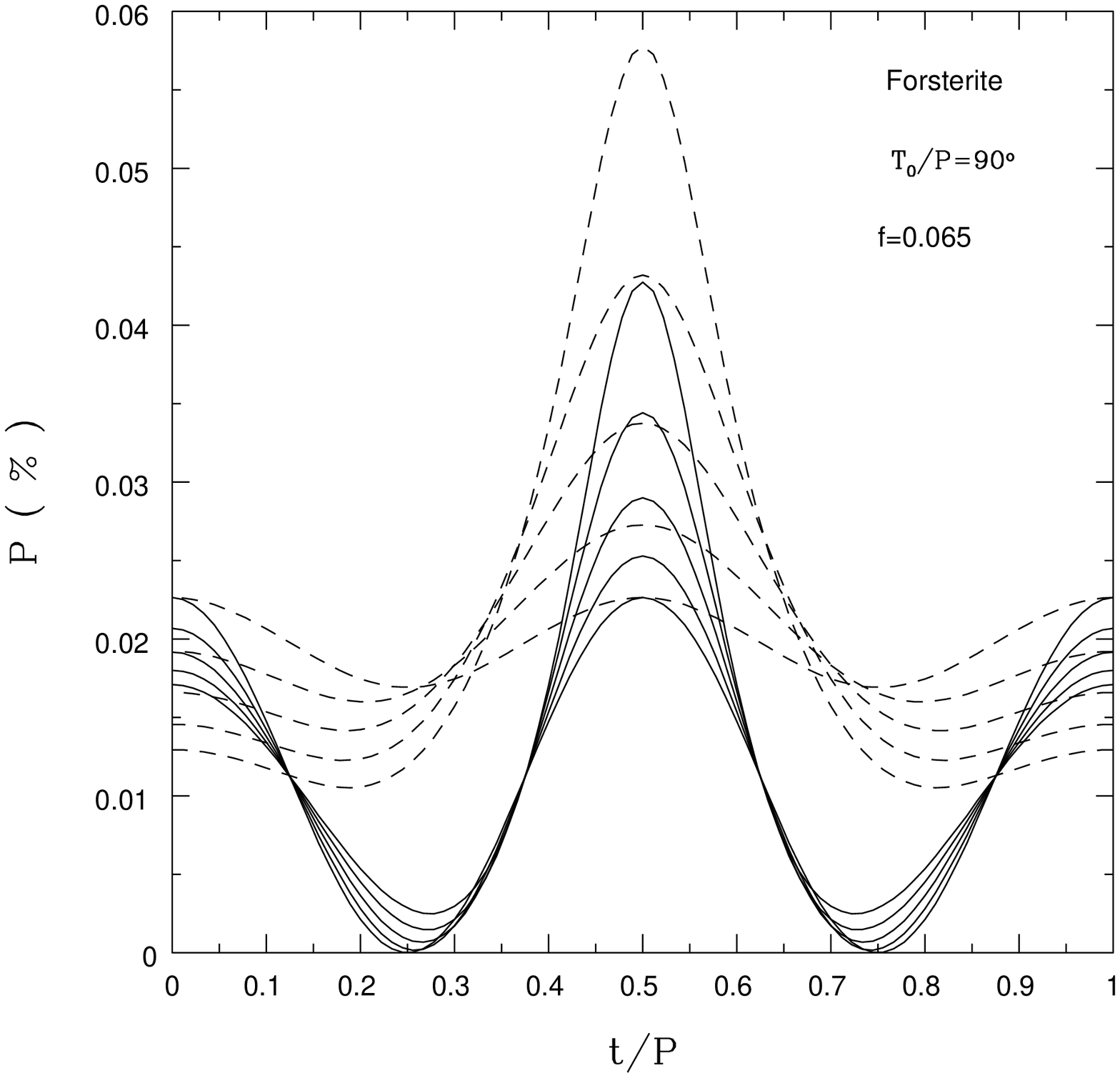}
\figcaption[f5.ps]{Degree of polarization by planet with f=0.065 and 
$T_0/P=90^o$ but different orbital eccentricity. $t/P$ is the fractional
orbital period, $t$ being time and $P$ the orbital period. Solid lines from
bottom to top at $t/P=0.5$ represent models with $e=$0.0, 0.1, 0.2, 0.3, and
0.4 respectively and with $i=90^o$  while the dashed lines represent that with
$i=30^o$. \label{figure5}}

\plotone{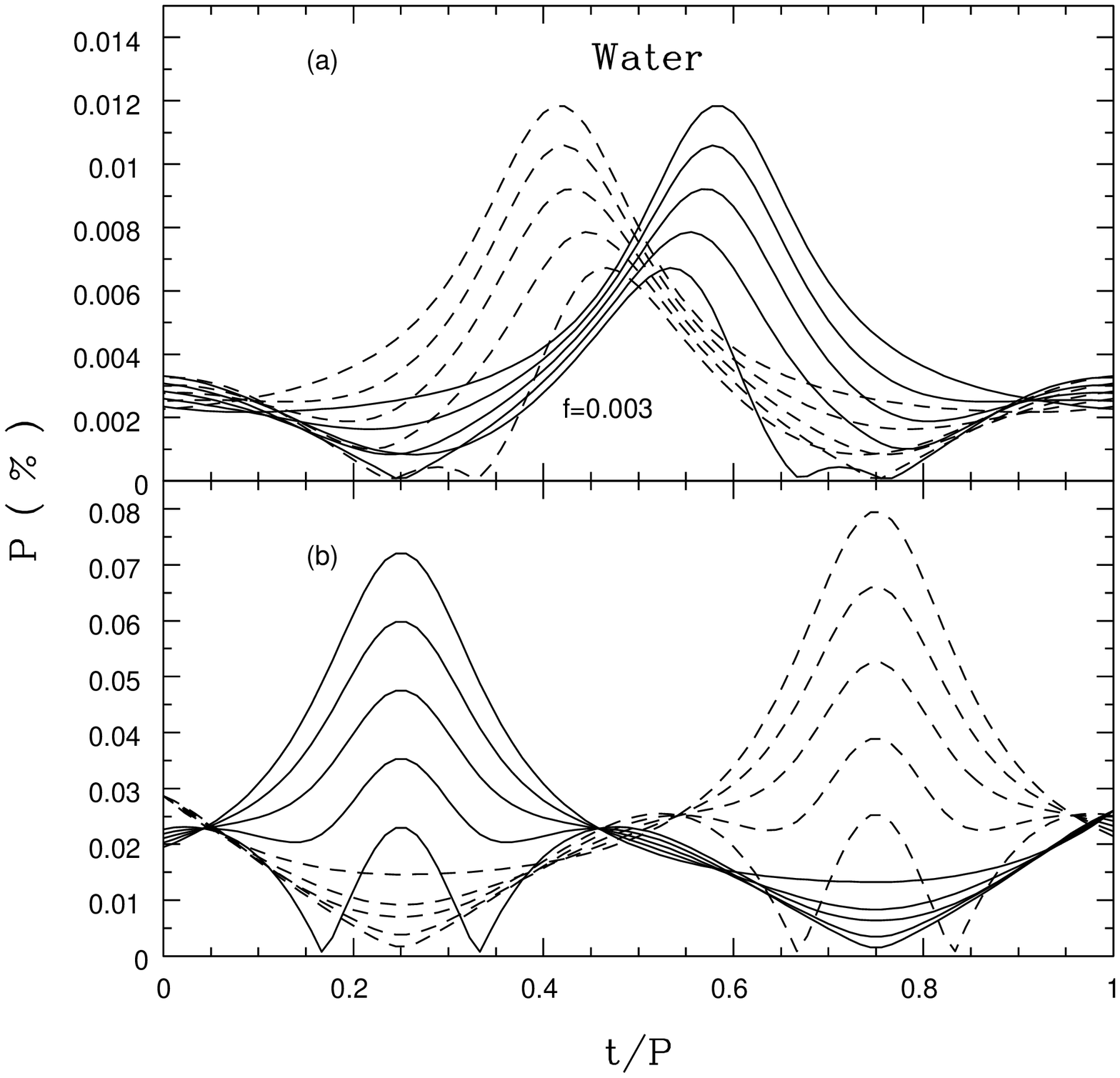}
\figcaption[f6.ps]{Degree of polarization as a function of fractional orbital
period $t/P$ at Bessel R band due to water condensates.  
(a) Models for planet with f=0.003, $e=0.4$ and mean particle
radius $r_0=5.0 \mu m$. Solid lines - models with $\Lambda_P=60^o$
($T_0/P=0.67$); dashed lines -  models with $\Lambda_P=120^o$
($T_0/P=0.42$).  Both the solid and the 
dashed lines from bottom to top at $t/P=0.25$ represent polarization with 
$i=90^o$, $60^o$, $45^o$, $30^o$ and $0^o$  respectively.
(b) Same as (a) but solid lines -  models with $r_0=1.0 \mu m$, 
f=0.001 and $T_0/P=0.25$; dashed lines -  models with
$r_0=4.0 \mu m$, f=0.01 and $T_0/P=0.75$.  \label{figure6}}

\plotone{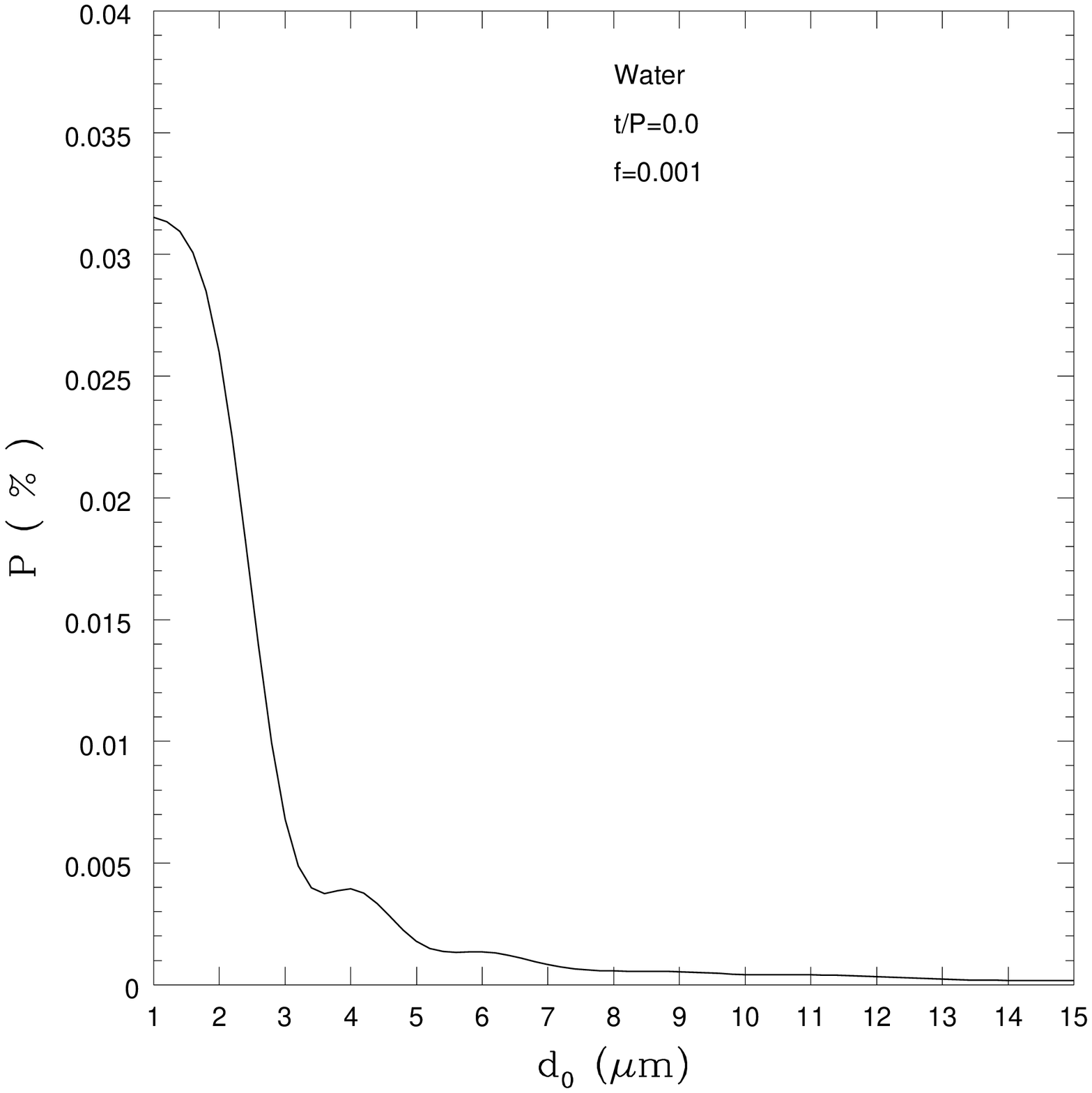}
\figcaption[f7.ps]{Degree of polarization at Bessel R band
as a function of mean grain diameter $d_0$.  Circular orbit with inclination
angle $i=90^o$.  The linear polarization by planet with oblateness f=0.001
having water is calculated at $t/P=0.0$.
\label{figure7}}
\end{document}